\documentclass[prb ,twocolumn,superscriptaddress]{revtex4}

\usepackage{graphicx}
\usepackage{dcolumn}
\usepackage{bm}
\usepackage{amsmath,amssymb}
\newcommand{\be}{\begin{equation}}
\newcommand{\ee}{\end{equation}}
\newcommand{\beq}{\begin{eqnarray}}
\newcommand{\eeq}{\end{eqnarray}}
\begin{document}

\title{Interferometric probe of paired states}

\author{V.~Gritsev},
\author{E.~Demler}
\affiliation{Department of Physics, Harvard University, 17 Oxford
Street, Cambridge, MA 02138}
\author{A.~Polkovnikov}
\affiliation{Department of Physics, Boston University, Boston, MA
02215}

\begin{abstract}
We propose a new method for detecting paired states in either
bosonic or fermionic systems using interference experiments with
independent or weakly coupled low dimensional systems. We
demonstrate that our method can be used to detect both the FFLO and
the d-wave paired states of fermions, as well as quasicondensates of
singlet pairs for polar F=1 atoms in two dimensional systems. We
discuss how this method can be used to perform phase-sensitive
determination of the symmetry of the pairing amplitude.
\end{abstract}

\date{\today}

\maketitle

\section{Introduction}


Interference experiments are the primary tool of detecting and
characterizing cold atom systems \cite{berman,schmiedmayer}. While
original experiments focused on demonstrating macroscopic coherence
of large BEC's~[\onlinecite{mit}], subsequent work used interference
experiments to explore more interesting phases and phenomena. For
example, interference in the time of flight (TOF) experiments was
used for observation of the superfluid to Mott insulator transition
in optical lattices~\cite{greiner}, analysis of fluctuations in low
dimensional systems~\cite{paris,heidelberg}, and studies of phase
diffusion and decoherence in dynamically split
condensates~\cite{mit,vienna,nist}. Interference can also give rise
to interesting patterns in second order coherence \cite{adl}. This
approach was used to demonstrate  that Hanburry Brown Twiss
experiments with both bosons and fermions~\cite{bloch1, bloch2,
ottl, NIST, schel, jeltes, Flor, jin1} and to observe pairing of
fermions\cite{jin1}. A series of recent theoretical and experimental
papers explored the idea that one can use interference between two
or more low-dimensional systems to probe their non-trivial
correlation functions~\cite{paris, heidelberg, PAD}. Partially using
this ideas Hadzibabic et. al. were able to detect
Berezinskii-Kosterlits-Thouless transition in two-dimensional
bosonic systems which is associated with vortex
proliferation~\cite{paris}. One of the novel feature of this
approach was the idea to use not only the average contrast but the
full distribution functions~\cite{GADP,IGD,heidelberg}. Distribution
functions are determined by high order correlation functions and
contain a wealth of information about underlying systems.
Distribution functions of interference fringe amplitudes were
recently analyzed for one-dimensional quasi-condensates and provided
direct probe of long wavelength phase fluctuations in the system of
either quantum or thermal origin~[\onlinecite{heidelberg}]. However
there is another source of fluctuations of the fringe amplitude
which is purely quantum in nature. Namely, this is shot noise coming
from the discreteness of particles~\endnote{We emphasize that in the
wave picture discreteness of particles is purely quantum effect
coming from the number phase uncertainty~\cite{IGD,P}}. Shot noise
is especially strong in systems with short range single particle
correlations, in particular in fermionic systems. Thus
interferometric probes in such systems are intrinsically more
difficult than in the systems with long or quasi long range order
for which shot noise is less important than the low wavelength
thermal and/or quantum fluctuations~\cite{IGD,P}. Our emphasis on
low-dimensional systems has two main reasons: they exhibit exotic
phases more often and it is easy to perform interference experiments
with them. In this paper we focus on fermionic and bosonic systems
in low dimensions and continue to study the possibility of using
interferometry to probe strongly-correlated many-body states.

In interacting systems one is often interested in states which do
not have coherence of individual particles but exhibit a coherence
(or slowly decaying correlations) of particle pairs.  For example,
fermionic paired states are characterized by the pairing amplitude
\begin{eqnarray}
\Delta(r) = \int d\eta \, f(\eta) \,
c_\downarrow (r-\eta/2) \, c_\uparrow (r+\eta/2)
\label{Delta_eq}
\end{eqnarray}
Here $r$ is the center of mass position of Cooper pairs and
$f(\eta)$ is the Cooper pair wave function~\cite{textbook}.  The
ordered state corresponds to the condensation of pairs of particles
and should be analyzed using correlation functions of the form
$\langle \Delta^\dagger(r_1) \Delta(r_2) \rangle$.  Correlation
functions of this type which we will refer to as anomalous
correlation functions also arise in the context of exotic states of
interacting bosons such as condensates of pairs of
bosons\cite{kuklov} and polar condensates in two dimensional
systems\cite{mukerjee,podolsky}. In principle one can extract
anomalous correlation functions analyzing higher order moments of
the interference amplitude. However, as we will show below, this
might be a very difficult task in practice because of effects of
shot noise \cite{IGD, P} and because such anomalous correlation
functions can appear as small corrections on top of normal
correlation functions.

In this paper we suggest an alternative method for identifying
paired states and for measuring directly their anomalous correlation
functions using interference experiments with two (or more) systems.
This paper extends earlier work on the analysis of interference
experiments with pairs of independent condensates of single
component bosons~\cite{PAD,GADP,IGD,paris,heidelberg}. Our main
purpose here is to show that one can probe fermionic superfluidity
in low dimensional systems. In particular, we define a new
observable, which we refer to in the text as anomalous interference
amplitude, which should vanish when there is no pairing between
fermions and which is nonzero when there is paring in the system. We
suggest two methods to detect this anomalous amplitude. The first
approach relies on detecting interferometric signal in two disjoint
parts of the system $R_{I,II}$ and averaging appropriate observable
over these disjoint regions. This way of detecting pairing
correlations relies on the existence of the long range (or quasi
long range) order in the pairing channel and thus requires phase
ordering in the fermionic superfluids. Note that averaging over two
disjoint regions is necessary to cancel the effects of an undefined
relative phase of the superfluid order parameter in two independent
layers. One can straightforwardly extend this idea and split the
system to a larger number of disjoint regions improving the signal
to noise ratio but other than that not affecting our analysis. In
the second method we introduce a weak tunneling coupling between the
systems to lock the relative phase. We show that in sufficiently
large systems there is always a broad range of parameters, where the
coherence is established but the correlation functions are still not
affected by the presence of this weak tunneling term. Because the phase
locking transition does not require long range order in each
superfluid, this method is more sensitive to the formation of the
local pairing amplitude. We further argue that in lattice fermionic
systems one can measure the symmetry of the pairing gap and thus
distinguish, for example, $d-$ wave from $s-$ wave superfluidity.
This can be achieved by aligning the probing laser beam along
different axes of the lattice.

The ideas presented in this paper can be further extended to
low-dimensional Bose systems. We show that in a similar setup one
can measure anomalous correlators in {\it bosonic} superfluids.
These correlators have an unusual property that they grow with the
separation between the particles showing effective ``anti-bunching''
behavior for bosons. Usually anomalous correlations are not easy to
detect, since they are not {\it gauge invariant}, i.e. they are
sensitive to the global superfluid phase. The two setups considered
here eliminate effects of this phase and make such measurements
possible.

Carusotto and Castin have previously suggested an experiment which
relies on particle interference to detect paired
states\cite{carusotto}. While there is some conceptual connection
between their work and our approach, our method has an advantage
that it does not require Bragg out-coupling of atoms, splitting and
mixing of atom beams, and using single atom detectors to measure
coincidences. As we demonstrate below, interference of two
ballistically expanding independent clouds does all of this work
itself!


The paper is organized as follows. In Sec.~\ref{sec:1} we first
analyze the basic structure of anomalous correlators and the
interference amplitude between two independent fermionic
superfluids. We then introduce the new observable, the anomalous
interference amplitude, which probes the pairing amplitude. In
Sec.~\ref{sec:schemeI} we show how this anomalous amplitude can be
detected performing simultaneous measurements in disjoint parts of
the time of flight image. Using this scheme we discuss possible
set-ups for observing the d-wave superfluid and the FFLO phases. We
suggest how one can detect not only the amplitude, but also a phase
of the pairing function. We perform explicit quantitative analysis
of the anomalous amplitude for two-dimensional superfluids with $s-$
and $d-$ wave pairing based on BCS-theory.
Then in Sec.~\ref{sec:anom} we discuss the second way of detecting
anomalous interference amplitude by introducing a weak interlayer
tunneling. We show that on the one hand its presence introduces
corrections to the results of Sec.~\ref{sec:1}, which are not
related to the superfluidity. On the other hand the presence of this
tunneling establishes the interlayer phase coherence. We show that
by decreasing the imaging area and increasing the system size one
can always achieve the regime where the coherence between the
superfluids is established and yet the effect of the tunneling on
the correlation functions is negligible. In Sec.~\ref{sec:3} we
extend our analysis to bosonic superfluids. In particular, we show
that in the superfluids with quasi long range order the anomalous
interference amplitude grows superlinearly with the imaging size
$\mathcal A$. In turn this implies that the corresponding
interference contrast increases with $\mathcal A$. This behavior is
opposite to that of the normal interference amplitude, which always
decreases with $\mathcal A$. And finally in Sec.~\ref{sec:4} we
summarize our results.

Throughout the paper we use BCS approximation to perform explicit
calculations. This approximation is only reliable in the weak
coupling regime; at strong coupling one has to do more elaborate
calculations. However, we do not expect any qualitative difference
between BCS and exact results.

\section{Analysis of the interference amplitude: Basic set-up}
\label{sec:1}
We start our discussion from analyzing the interference amplitude of
two fermionic condensates. Extension of our results to the case of a
stack of several condensates is straightforward. For concreteness we
will focus on the case of two dimensions. First we analyze the usual
interference amplitude, which is determined by normal correlation
functions and show that it is not a reliable detection tool of
superfluidity. Then we describe how one can use the same
interference experiments but analyze the results differently to
extract {\it anomalous} correlation functions.

\subsection{Normal correlation functions.}
\label{sec:norm}

Consider two independent systems (layers) and assume that each
system contains two species of atoms, which we label by a spin index
$\sigma$. Let $c^\dagger_{i\sigma}({\bf r})$ be the creation
operators for atoms with spin $\sigma$ in layer $i=1,2$ and the in
plane coordinate ${\bf r}$. After the expansion we find interference
fringes in the $z$ direction, so that the density $\rho_{{\rm
int},\, \sigma}(z,{\bf r})\sim C_\sigma({\bf r}) \cos ( Q z +
\phi_\sigma({\bf r}) )$ where $Q=md/\hbar t$ (this assumes
sufficiently long expansion time, see e.g. Ref.~[\onlinecite{BDZ}]).
Because the phase $\phi_\sigma({\bf r})$ is a random variable for
independent systems the average density does not show any
interference fringes. Thus to filter out this oscillating component
we have to consider Fourier transform of the density-density
correlation function. Indeed one can chose the following operator,
which corresponds to the square of the interference
amplitude~\cite{P}:
\beq
|A|^{2}&=&\pm\int \rho(z_{1},{\bf r}_{1},t)\rho(z_{2},{\bf r}_{2},
t)e^{iQ(z_{1}-z_{2})}\nonumber\\&\times&dz_{1}dz_{2}d{\bf
r}_{1}d{\bf r}_{2}\mp\int \rho(z,{\bf r},t) dz d{\bf r}
\label{A_Q0},
\eeq
where the upper (lower) sign corresponds to bosons (fermions). Here
$\rho(z,{\bf r},t)$ is the atomic density at position $z,{\bf r}$ at
time $t$ after the expansion. The $z$ coordinate is orthogonal to
the atomic systems, while ${\bf r}$ describes positions of the atoms
within each individual system. For two dimensional systems,
integration over one of the directions is done automatically by the
laser beam, whereas integration in the other direction is done
manually\cite{paris}.  We assume that the transverse confinement is
tight and when the atoms are released, they expand strongly in the
transverse $z$ direction, while their in plane expansion can be
neglected. This assumption is well justified if the transverse
confining energy is large compared to any other energy scales in the
problem.

Before explaining where the expression (\ref{A_Q0}) came from let us
investigate it a little further. Assuming that the long time of
flight allows us to use the far field expressions~\cite{IGD,P,zoran}
we find
\be\label{A_Q}
|A|^{2}
=\sum_{\sigma,\sigma'}\int \int d{\bf r_{1}}d{\bf r_{2}}
c_{1,\sigma}^\dagger({\bf r}_{1})
c_{2,\sigma^\prime}^\dagger({\bf r_{2}})c_{2,\sigma^\prime}({\bf r}_{1}) c_{1,\sigma}({\bf r}_{2}).
\ee
Note that both for bosons and the fermions the expression above can
be obtained from the complex interference amplitude defined as
\be
A=\sum_\sigma
\int d{\bf r} \, A_{\sigma}({\bf r}) = \int d{\bf r}
\, c^\dagger_{1\sigma}({\bf r}) c_{2\sigma}({\bf r}).
\label{Asigma}
\ee
On can think about $A$ as of the Fourier transform of the density of
the expanded cloud in the $z$ direction (for a given ${\bf
r}$)\endnote{Interactions during the initial moments of expansion
result in finite momentum broadening of the Fourier transform around
$Q=md/\hbar t$. This effect can be suppressed by increasing
transverse confinement of atoms.}. Then the expression (\ref{A_Q})
can be obtained as the normal ordered product of $A^\dagger A$:
\be
|A|^2=\pm :A^\dagger A:,
\ee
where the ``$+$'' sign corresponds to bosons and the ``$-$'' sign does to fermions.

In the case of independent systems there is no coherence between
atoms hence expectation value of $A$ is zero. This does not mean the
absence of interference fringes in individual shots but only tells
us about the random phase of interference fringes. Indeed the
quantity $|A|^2$ is insensitive to this phase. It directly measures
the (square of the) amplitude of the interference and it does not
average to zero even for independent systems.

Let us make a few comments on where the expressions above come from.
In the Eq.~(\ref{A_Q0}) we are taking Fourier transform of the
product of the densities of atoms after expansion. This Fourier
transform picks the component in this product oscillating with the
wavevector $Q$ and thus corresponding to the interference between
the two systems. Note that the operator $|A|^2$  in Eq.~(\ref{A_Q0})
is very similar to the one originally introduced for
bosons~\cite{PAD} except for the negative sign appearing for
fermions and except for the additional second term. The negative
sign takes care of the fermionic statistics, or equivalently of the
additional $\pi$ phase shift in the interference part of the
density-density correlation functions~\cite{P}. The second term in
Eq.~(\ref{A_Q0}) removes the trivial contribution to the Fourier
transform coming from shot noise which is not related to the
interference. This term is usually unimportant for bosonic systems.
Note that Eq.~(\ref{A_Q0}) can be rewritten using the normal ordered
product of densities:
\be
|A|^{2}=\int :\rho(z_{1},{\bf r}_{1},t)\rho(z_{2},{\bf r}_{2},
t):e^{iQ(z_{1}-z_{2})}.
\label{A_Q1}
\ee
Substituting the far field expansion of the bosonic
operators~\cite{IGD, P, zoran} into Eq.~(\ref{A_Q1}) we easily
recover Eq.~(\ref{A_Q}). We emphasize that it is important to first
take the normal order in the product of densities $\rho(z_{1},{\bf
r}_{1},t)\rho(z_{2},{\bf r}_{2},t)$ and only after use the far field
expansion for the density operators. Using the opposite order will
give spurious contributions. In bosonic systems with large number of
atoms in the same state the creation and annihilation operators can
be approximately treated as commuting classical fields and thus no
ambiguity with ordering appears and the shot noise contribution is
small~\cite{IGD, P}. However, for fermionic systems, where the shot
noise is usually important one has to be careful in evaluating
integrals like those appearing in Eq.~(\ref{A_Q0}).

For bosons statistical and scaling properties of $|A|^2$ contain
important information about superfluidity~\cite{PAD, GADP}, which
can be straightforwardly detected in experiments~\cite{paris,
vienna}. At the same time for fermions information about
superfluidity is encoded in the Cooper pair correlation functions.
Although the operator $|A|^2$ certainly contains the information
about superfluidity (see Appendix~\ref{App:A}) and can be in
principle used to determine the pairing, it does not provide a
"smoking gun" for detecting superfluidity. Indeed pairing only
quantitatively affects the magnitude of the interference amplitude
$|A|$. This magnitude can be affected also by various other reasons.
Thus it is important to find another observable which vanishes
unless fermions are paired. We are going to introduce such an
observable in the next section.

\subsection{Anomalous correlation functions}

An observable, which directly probes the pairing wave function can
be constructed from Eq.~(\ref{A_Q0}) with a slight modification:
\beq
&&A_Q^2=\int \rho(z_{1},{\bf r}_{1},t)\rho(z_{2},{\bf r}_{2},
t)\mathrm
e^{iQ(z_{1}+z_{2})}dz_{1}dz_{2} d{\bf r}_{1}d{\bf r}_{2}.\nonumber\\
\label{B_Q0}
\eeq
Note the difference between Eqs.~(\ref{A_Q0}) and (\ref{B_Q0}). The
former corresponds to taking the product of the Fourier transform of
the density $\rho(z)$ and its complex conjugate. The latter
corresponds to taking the square of the Fourier amplitude without
taking complex conjugation. For the long expansion time this
expression reduces to
\beq
A^2 &=&\sum_{\sigma,\sigma'}\int\int d{\bf r}_{1} d{\bf r}_{2}
c_{1,\sigma}^\dagger ({\bf r}_{1})c_{1,\sigma^\prime}^\dagger ({\bf
r}_{2})\nonumber\\
&\times& c_{2,\sigma^\prime}({\bf r}_{2})c_{2,\sigma}({\bf r}_{1}).
\label{B_Q}
\eeq

The quantity $A^2\equiv :A^{2}:$ looks like exactly what we need.
Indeed for independent layers it depends only on the product of the
pairing amplitudes in the two layers:
\begin{eqnarray}
A_2\equiv\langle A^{2}\rangle = \int\int d{\bf r}_1 d{\bf r}_2 \,
F_1^\star({\bf r}_1,{\bf r}_2) F_2({\bf r}_1,{\bf r}_2),
\label{Eq:BQ2}
\end{eqnarray}
where $F_1^\star({\bf r}_1,{\bf r}_2) = \langle
c^\dagger_{1\uparrow}({\bf r}_1) c^\dagger_{1\downarrow}({\bf
r}_2)\rangle$. However, there is one subtlety. Unlike the normal
amplitude squared $|A|^2$, which is always a positive real number,
the anomalous amplitude squared $A^2$ is complex. Moreover for
independent condensates $A^2$ is equal to zero because the phases of
$F_1$ and $F_2$ are not correlated. To avoid this phase uncertainty
one can try to look into $|A^2|^2$, which will involve second order
correlation functions in each layer. However, it is easy to see that
$|A^2|^2$ will be dominated by shot noise and normal (not anomalous)
correlation functions. Thus there will be no advantage compared to
analyzing $A^2$.

The main purpose of this paper is to show that one can overcome the
effect of the uncertain relative phase and measure the anomalous
amplitude $A$ and thus detect superfluidity in fermionic systems. We
note that fundamental reason why extra efforts are needed to measure
$A^2$ compared to $A^2$ is because the former (for independent
systems) is not a {\it gauge invariant quantity}. This difficulty
will similarly arise if one tries to measure not gauge invariant
quantities in other setups. The ideas of this work can be extended
to those situations as well. In later sections we will discuss some
other examples of this kind.

\begin{figure}[ht]
\includegraphics[width=8cm]{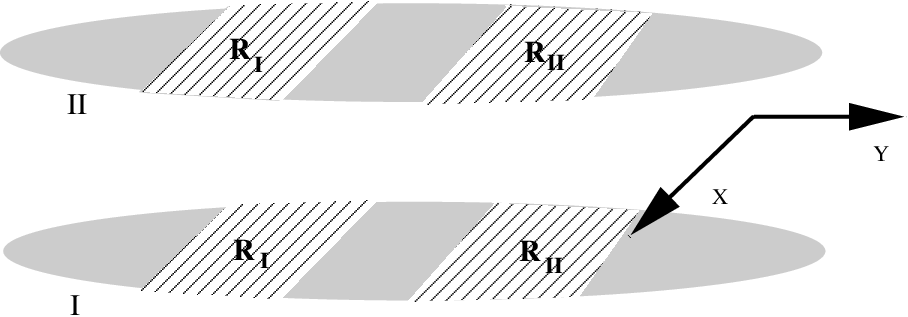}
\caption{The experimental set-up corresponding to the Scheme I of
our approach. The integration in $X$-direction is performed
automatically by the imagine beam, whereas the integration in the
$Y$-direction is done by "hands". The interference signal is
collected from two disjoint regions $R_{I}$ and $R_{II}$.}
\label{scheme1}
\end{figure}

Here we suggest two different setups to fix the problem with the
unknown phase in Eq.~(\ref{Eq:BQ2}). In the first setup, which we
refer to as Scheme I (see Fig.~(\ref{scheme1})), we get rid of the
random phase by making a special choice of the spatially separated
integration domains. Namely instead of integrating $A^2$ over the
entire region one splits the imaging area spanned by ${\bf r_1}$ and
${\bf r}_2$ to two spatially separated domains $R_I$ and $R_{II}$.
In each experimental run one independently determines $A^2$ in the
two domains then takes the absolute value of their square and
averages over many experimental runs. As we will show in detail this
setup relies on the fact that single particle correlation functions
decay sufficiently fast with the distance, while the pair
correlation functions decay slowly or do not decay at all. This
setup has an obvious advantage compared to measuring $|A^2|^2$
because single particle normal correlation functions decay fast with
the distance. As a result the quantity $A^2(\Omega_1)A^{\star\,
2}(\Omega_2)$ is dominated by anomalous correlation functions:
\begin{displaymath}
\langle c_{\uparrow,I}^\dagger c_{\downarrow,I}^\dagger
c_{\downarrow,II} c_{\uparrow,II}\rangle\approx \langle
c_{\uparrow,I}^\dagger c_{\downarrow,I}^\dagger\rangle\langle
c_{\downarrow,II} c_ {\uparrow,II}\rangle,
\end{displaymath}
where subscripts $I$ and $II$ indicate that spatially these
operators are located in regions $R_I$ and $R_{II}$. Note that if
the domains are not spatially separated or single particle
correlations functions do not decay fast there are additional
(unwanted) cross correlations in the equation above like $\langle
c_{\uparrow,I}^\dagger c_{\uparrow,II}^\dagger\rangle \langle
c_{\downarrow,I}^\dagger c_{\downarrow,II}\rangle$.

\begin{figure}[ht]
\includegraphics[width=8cm]{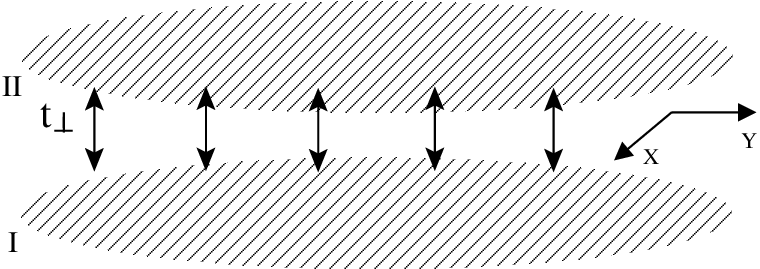}
\caption{The experimental set-up corresponding to the Scheme II of
our approach. The non-zero tunneling between two layers locks the
phases of superfluids. The interference signal is collected now from
the whole areas of superfluids.}
\label{scheme2}
\end{figure}

In the second setup, which we refer as Scheme II, (see
Fig.~(\ref{scheme2})) we introduce a weak tunneling $t_{\perp}$
between the two layers. This tunneling locks the phases of pairing
amplitudes in the two layers and makes the expectation value of
$A^2$ real and positive. Besides the phase locking effect, the
tunneling induces the mixing between the fermions in the two layers
and results to the nonzero contribution to $A^2$ in Eq.~(\ref{B_Q})
even in the absence of pairing. Below we will show that at small
temperatures it is always possible to choose the tunneling such that
phase locking transition already occurred but the correlation
functions are not yet significantly affected so that
Eq.~(\ref{B_Q2}) still holds.
The two setups are complimentary to each other and can be used
depending on the situation. The Scheme I essentially relies on the
existence of the long (or quasi-long) range order in the pairing
amplitude. As we will show this scheme can be adopted to measuring
not only the existence of superfluidity but also to the symmetry of
the order parameter and even its phase. While the Scheme II is more
sensitive to the local pairing between fermions and less to the
existence of long range order in the superfluid phase. Such setup
can be used, for example, to measure pseudo gap phenomena. Scheme II
can also be used to determine the local symmetry of the pairing wave
function but not its phase.

\section{Scheme I: Basic set-up and Various examples\label{sec:schemeI}}

Keeping the analogy with analysis of normal correlations we
emphasize the integration region $R$ in the definition of the
operator $A^{2}$ and denote it as $A^{2}(R)$ in what follows
\begin{eqnarray}
&&A^{2}(R_I) = A_{\uparrow}( R_I) A_{\downarrow}(R_I) = \nonumber\\
&=& \int_ {R_I} d{\bf r}_1 \, \int_ {R_I} d{\bf r}_2 \,
c^\dagger_{1\uparrow}({\bf r}_1) c^\dagger_{1\downarrow}({\bf r}_2)
c_{2\downarrow}({\bf r}_2) c_{2\uparrow}({\bf r}_1)
\label{B_Q2a}
\end{eqnarray}
Since for independent condensates the correlation function above
factorizes into a product of anomalous correlation functions in each
system and from  Eq.~(\ref{B_Q2a}) we arrive to Eq.~(\ref{Eq:BQ2}).
As we argued earlier, because $A^{2}(R_{I})$ is a complex number
with a phase which is random from shot to shot, taking expectation
value of equation (\ref{Eq:BQ2}) gives us zero. To get rid of this
random phase we compare interference patterns from a pair of
regions, $R_I$ and $R_{II}$. More precisely we take
$[A^{2}(R_I)]^\dagger A^{2}(R_{II})$, so that the random relative
phase between pairing functions $F_1$ and $F_2$ drops out.
Experimentally this procedure corresponds to taking a square of the
Fourier transform of the density along $x$-direction integrated over
the region $I$ and multiplying it by a complex conjugate of a
similar quantity integrated over the region two. The result of this
manipulation is then averaged over many experimental runs. Assuming
that the system has a true long range order, taking regions $R_I$
and $R_{II}$ to be separated by a distance which is appreciably
larger than the size of the Cooper pairs, and taking $R_I$ and
$R_{II}$ to be identical we find
\beq
&&\langle [A^{2}(R_I)]^\dagger A^{2}(R_{II}) \rangle =
\nonumber\\
&& \left| \int_{R_I} d{\bf r}_1 \, \int_{R_I} d{\bf r}_2 \, F_1^\star({\bf r}_1,{\bf r}_2) F_2({\bf r}_1,{\bf r}_2)\right|^2
\label{A_composite}
\eeq

We now consider several specific examples in which  analysis of
expectation values of the type (\ref{A_composite}) can be used to
identify interesting many-body states.

\subsection{Analysis of the anomalous interference  amplitude
 within the BCS theory}

We now consider the integral in Eq.~(\ref{A_composite}). To simplify
calculations we also assume the translational invariance in both
systems. In this case the pairing wave function depends only in the
difference between ${\bf r}_1$ and ${\bf r}_2$: $F_{\alpha}({\bf
r}_1,{\bf r}_2)\equiv F_{\alpha}({\bf r}_1-{\bf r}_2)$. Then
\be
A_2=2\mathcal A \int_{R} d{\bf r} F_1^\dagger({\bf r})F_2({\bf r}),
\label{B_Q2}
\ee
where the integration is again taken over the part of a condensate
$R_I$ or $R_{II}$ with the imaging area $\mathcal A$. As we noted
once the relative phase is taken care of and assuming the two
condensates are identical we have $F_1=F_2$ and thus
\be
A_2=2\mathcal A\int {d{\bf k}\over (2\pi)^2} |F({\bf k})|^2,
\label{B_Q3}
\ee
The integral above can be easily evaluated within the BCS model (we
take zero temperature limit)
\be
A_2= {\mathcal A\over 2}\int {d{\bf k}\over (2\pi)^2}{\Delta^2_{\bf
k}\over E_{\bf k}^2}.
\label{B_Q4}
\ee
where the pairing function $\Delta_{\bf k}$ has to be specified for
concrete type of pairing, $E_{\bf k}=\sqrt{\xi_{\bf
k}^{2}+\Delta_{\bf k}^{2}}$ and $\xi_{\bf k}={\bf k}^{2}/2m -\mu$ is
a single-particle dispersion. Thus if the pairing gap is isotropic
and energy independent $\Delta_{\bf k}\equiv \Delta_{0}$ then in 2D
we find
\be
A_2^{(s)}= {\mathcal A\rho_0\Delta_{0}\over 2}\arctan
{\Delta_{0}\over\mu}.
\label{B_Q5}
\ee
where the 2D constant density of states $\rho_{0}$ is introduced. In
the weak coupling limit $\Delta_{0}\ll\mu$, where $\mu$ is a
chemical potential, we have $A_2\approx N{\Delta^{2}_{0}/4\mu^2}$,
where $N\approx 2\mu\rho_0$ is the total number of particles (the
factor of two takes into account two different spin components). If
the pairing is strong then BCS extrapolation gives $A_2\approx
N\pi/4$. We see that $A_2$ is a monotonically increasing function of
the pairing gap and thus can serve as a direct probe of the
fermionic superfluidity. Note that Eq.~(\ref{B_Q4}) can be also
analyzed in the case of d-wave pairing, where $\Delta({\bf k})\sim
\Delta_0 \cos(2\theta_{\bf k})$, where $\theta_{\bf k}$ is the polar
angle of the wave vector ${\bf k}$. The result is (see Appendix C)
\be
A_2^{(d)}= {\mathcal A\rho_0\over
2}2\pi\mu\left(\sqrt{1+\left(\frac{\Delta_{0}}{\mu}\right)^{2}}-1\right).
\label{B_Q6}
\ee
However, since only the square of $\Delta_{\bf k}$ enters
Eq.~(\ref{B_Q4}) and we are explicitly averaging over angles, the
difference between $s-$ and $d-$ parings will be minor. In fact one
can show that in the $d-$wave case Eq.~(\ref{B_Q5}) gets multiplied
by a smooth function of $\Delta_0/\mu$ which changes between $1/2$
at $\Delta_0\ll \mu$ and $2/\pi$ in the opposite limit.

\subsection{Phase sensitive detection of the d-wave pairing}
\label{sec:phase}


In the section above we discussed a possibility to detect anisotropy
of the pairing amplitude using one-dimensional integration. In
particular for the d-wave paring (dSF) the interference signal
should vanish along the nodal directions. There are also other
earlier suggestions for the detection of dSF, which rely rely on the
detection of the Dirac like dispersion of
quasiaprticles~\cite{altman,georges,giamarchi_kohl,HCZDL}. This
however is not a unique signature of the d-wave pairing state. A
Dirac cone of quasi-particles may also arise for an anisotropic
s-wave pairing~\cite{scalapino} or d-density wave
states~\cite{chakravarty}.

Here we would like to show how the Scheme I can be extended to do
phase sensitive detection of dSF.  In high temperature cuprate
superconductors, the crucial experiments which identified the
$d$-wave character of pairing were phase sensitive experiments by
Van Harlingen et al.~\cite{vanHarlingen} and Tsuei and
Kirtley~\cite{TK}. These experiments unambiguously demonstrated the
correct angular dependence of the pairing amplitude. Experimental
set-up by Van Harlingen et al. used a combination of an s-wave and
d-wave superconductors in a corner SQUID geometry. Interference of
s-wave Cooper pairs with different parts of d-wave Cooper pairs was
used to establish the relative phase of of the Cooper pair wave
function.

What we discuss below is the cold atoms analogue of the Van
Harlingen experiments. Hence we also need a source of s-wave Cooper
pairs and a source of d-wave Cooper pairs. We imagine a pair of two
dimensional fermionic systems, made of the same species of atoms,
but having s-wave pairing in one layer and d-wave pairing in the
other layer. This may be achieved, for example, using magnetic field
dependence of the scattering length and applying a strong field
gradient. Now we analyze interference patterns from two regions,
$R_{I}$ and $R_{II}$, which differ only by the 90$^o$ rotation. The
quantity $A_Q(R_I)$ is a complex number which has a random phase
from one shot to another. Analogously $A_Q(R_{II})$ is a complex
number with a random phase. But the $d$-wave symmetry of the pairing
requires that phases of these two complex amplitudes differ by
precisely $\pi$. Hence one can look at $\langle [A_Q(R_I)]^\dagger
A_Q(R_{II}) \rangle $ and the d-wave symmetry dictates that this
expectation value should be negative. On the other hand, when
$R_{I}$ and $R_{II}$ have the same orientation, expectation value of
$\langle[ A_Q(R_I)]^\dagger A_Q(R_{II}) \rangle $ should be a
positive number. It is important to emphasize that this statement is
general and does not rely on the specific microscopic model for
$d$-wave pairing. We stress that that only one of the layers should
have a d-wave symmetry otherwise $A_Q$, being proportional to the
product of anomalous correlation functions in two layers, does not
change sign under rotations (see Eq.~(\ref{B_Q2})). While the
precise value of $\langle[A_Q(R_I)]^\dagger A_Q(R_{II})\rangle $ is
not easy to calculate, especially if we are dealing with
non-identical superfluids, that statement of the $\pi$ phase
difference between $A_Q(R_I)$ and $B_2(R_{II})$ relies only on the
$d$-wave nature of pairing.

The crucial feature of the method discussed in this subsection is
that it should provide a qualitative and model independent
signatures of d-wave pairing. It does not rely on detailed analysis
of the microscopic models but it uses only the fundamental symmetry
of the d-wave order parameter.

\subsection{Probing of the anisotropy of pairing amplitude\label{probing-pairing}}

We now discuss another probe of d-wave pairing. Unlike the previous
method, it can not be used to demonstrate the change of the sign of
the gap function $\Delta(k)$. However it can be used to observe
anisotropy of the gap. We study the correlation function (\ref{B_Q})
but integrate it in a highly anisotropic way. In particular, one
length, say along the probing beam should be macroscopic and the
other $W$ should be shorter than the coherence length. Then (the
square of) the anomalous interference amplitude becomes
\be
a_2(\theta)=2\mathcal A W \int dz |F({z,\theta})|^2,
\label{b_Q}
\ee
where $\theta$ is the polar angle which defines direction of
integration. We introduced a new notation $a_2$ to avoid possible
confusion with $A_2$ analyzed earlier. Note that typically d-wave
symmetry of the order parameter requires the presence of the optical
lattice. This lattice in turn breaks rotational symmetry in the
superfluid and locks the phase of the pairing amplitude with the
lattice's principal axes. Therefore there is no ambiguity in
defining $\theta$ from one experimental run to another. One can
expect that for $s-$ wave pairing (\ref{b_Q}) should give isotropic
result, while for the $d-$ wave pairing the outcome will be highly
anisotropic. While this approach does not provide a "smoking gun"
signature of the change of sign in the Cooper pair wavefunction,
this method is easier to do experimentally; if successful it should
provide a strong indication of anisotropic pairing.

It is straightforward to show that in the $s-$wave case the function
$F$ is isotropic and is given by
\be
F(z,\theta)=\rho_0\Delta\, J_0(z/\zeta)K_0(z/\xi),
\label{Fz}
\ee
where
\begin{displaymath}
\zeta^2={1\over \sqrt{k_\mu^4+k_\Delta^4}+k_\mu^2},\quad
\xi^2={1\over \sqrt{k_\mu^4+k_\Delta^4}-k_\mu^2},
\end{displaymath}
$k_\mu=\sqrt{2m\mu/\hbar^2}$ and $k_\Delta=\sqrt{2m\Delta/\hbar^2}$.
This expression shows that the pairing wavefunction diverges
logarithmically at small $z$ and decays exponentially with the
characteristic correlation length $\xi$ at large $z$. The
logarithmic divergence is the usual artifact of the BCS theory with
point-like interactions. This divergence is cutoff at short
distances.

Using Eqs.~(\ref{Fz}) we evaluate the integral in Eq.~(\ref{B_Q6})
and find
\beq
&&a_2=\frac{\pi^2}{2} \mathcal A W\rho_0^2\Delta^2\xi\nonumber\\
&&~~~~~~~\times\!\!\!{\phantom x}_4F_3\left[\left({1\over 2},{1\over
2},{1\over 2},{1\over 2}\right),(1,1,1),-{\xi^2\over
\zeta^2}\right],
\eeq
where $_4F_3$ is the generalized Hypergeometric function. At small
and large ratio of $\Delta/\mu$ the expression above gives the
following asymptotics:
\beq
a_2 &\approx& {\sqrt{2}\over 3\pi} \mathcal A W
{\rho_0^2\Delta^2\over k_\mu}\ln^3\left(\Delta\over\mu\right),
\quad \Delta\ll\mu;\\
a_2 &\approx& \frac{18.8}{4}\, \mathcal A W{\rho_0^2\Delta^2\over
k_\Delta}, \qquad \Delta\gg\mu.
\eeq
As before the high energy estimate of the asymptotics is the
extrapolation of the BCS result to the strong coupling limit. For
$d-$ wave pairing the correlation function $F(z_\theta,0)$ vanishes
along the nodal direction and thus $a_{Q}(\theta)$ should vanish as
well. On the other hand, along the antinodal direction we can
recover the asymptotics similar to $s-$ wave case,
\beq\label{bQd}
a_2\approx \frac{1}{6\pi} \mathcal A W {\rho_0^2\Delta^2\over
k_\mu}\ln^3\left(\frac{\Delta}{8\sqrt{2}\mu})\right), \quad
\Delta\ll\mu.
\eeq
For details of computations see Appendices B and C.

\subsection{FFLO phase}

One of the most intriguing suggestions for the paired states of
fermions with attractive interactions is the idea of FFLO phase for
systems with spin imbalance. This state corresponds to Cooper
pairing at a finite momentum and has been a subject of extensive
theoretical studies during the last couple of
years\cite{combescot,yip,radzihovsky,huse,drummond,others}.
Experimental situation remains unclear (for recent review see
Ref.~[\onlinecite{KZ}]). We now discuss how interference experiments
can be adopted to look for the FFLO state. An earlier proposal for the
detection of the FFLO phase can be found in Ref.~[\onlinecite{kunyang}].

The FFLO phase is characterized by the finite center of mass
momentum of the Cooper pairs so that $F({\bf r}_1,{\bf r}_2)=
\langle c_{\uparrow}({\bf r}_1)c_{\downarrow} ({\bf
r}_2)\rangle\propto e^{\pm i {\bf Q} ({\bf r}_1+{\bf r}_2)/2}$.
Therefore when one analyzes the anomalous interference amplitude one
expects additional modulations which can be detected by taking an
appropriate Fourier transform. This can be achieved by changing the
integration procedure in Eqs.~(\ref{Asigma}) and (\ref{B_Q2}). We
note that this integration is not equivalent for the two directions.
In the direction of the $x$ axis, integration is done automatically
by the laser beam. In the other direction, i.e.  along the $y$ axis,
it is performed ``manually'' by integrating interference fringes
(see Fig.~\ref{scheme2}). An alternative approach is to take a
Fourier transform of the interference amplitude $A_\sigma(y)= \int
dx A_\sigma (x,y)$ along the $y$ axis. This can also be thought of
as modifying the integral in equation (\ref{Asigma})
\begin{eqnarray}
A_{Q,\sigma}(R) =\int_ {R}A_\sigma({\bf r})e^{iQ_{x}x}d{\bf r}=
\int_ {R} d{\bf r}e^{iQy}
c^\dagger_{1\sigma}({\bf r}) c_{2\sigma}({\bf r})\nonumber\\
\label{A_Q_sigma}
\end{eqnarray}
where we implicitly assume that the direction of the vector ${\bf
Q}$ coincides with the direction of integration $x$, ${\bf
Q}=(Q,0)$. Defining now
\begin{eqnarray}
&&A_2(R)=\langle A_{Q,\sigma}(R)A_{Q,\sigma'}(R) \rangle =
\nonumber\\ &=&\int_{R}d{\bf r}_1\int_{R}d{\bf r}_{2}
e^{iQ(y_1+y_2)} F_1^\star (r_1,r_2) F_{2}(r_1,r_2).
\end{eqnarray}
In the FFLO phase $A_2$ should have additional peaks at $Q$ matching
the finite momentum of the Cooper pair. The global unknown relative
phase can be removed again either by multiplying the signal coming
from two spatially separated imaging areas $R_I$ and $R_{II}$ or by
introducing weak tunneling coupling between the layers as discussed
in the next section.

One may be concerned that in rotationally invariant systems the
direction of the FFLO ordering wavevectors will not generically
coincide with the $x$ axis used for the observation. This issue should be
avoided by using systems that do not have a rotational symmetry
in the $xy$ plane. In fact, one of the most promising systems for
observing the FFLO phase is an array of weakly coupled 1d
systems~\cite{huse, drummond}. In this case the ordering wave vector should
be in the direction of the tubes.

\section{Scheme II: Anomalous Correlation Functions in
phase locked systems}
\label{sec:anom}

Another way to overcome the effect of unknown relative phase
detecting anomalous interference amplitude is to introduce a weak
tunneling between the two layers (see Fig.~\ref{scheme2}). As it was
shown in Refs.~[\onlinecite{mathey,giamarchi}] such tunneling leads
to the phase-locking transition. At the same time if the tunneling
is sufficiently weak then correlation functions do not appreciably
change and Eq.~(\ref{B_Q2}) is still valid. Below we will show that
there is indeed a wide range of parameters where the phases between
pairing amplitudes in two layers are locked and Eq.~(\ref{B_Q2})
gives the dominant contribution into the expression (\ref{B_Q}).

In the next section we analyze the effect of weak coupling more
carefully. We will explicitly analyze only the case of two coupled
$s$-wave super fluids. However, our results should be very general
because precise nature of the symmetry of the pairing amplitude
($s$-wave, $d$-wave, FFLO, etc.) is not very important for the
phase-locking phenomena.

\subsection{Role of the inter-layer coupling}
\label{sec:cor}

Two imaging areas used in the set-ups discussed previously were
needed to  cancel the unknown relative phase between the order
parameters in two layers. The same effect, however, can be achieved
by introducing a weak tunneling coupling between the layers. Then
phases of order parameters should lock~\cite{mathey, giamarchi} and
one does not have to combine signals from two different areas.
Conversely different imaging areas can be used as independent
sources so that one can effectively average $A_2$ over several
independent imaging areas in a single experimental run.

As before we will work in the BCS limit. The BCS Hamiltonian of two
coupled condensates reads
\beq
&&\mathcal H=\sum_{\bf k,\,\alpha} \psi_\alpha^\dagger({\bf
k})\left( \varepsilon_{\bf k}\tau_z+\Delta_{\bf
k}\tau_x\right)\psi_\alpha({\bf k})\nonumber\\
&& -t_\perp \sum_{\bf k}(\psi_1^\dagger({\bf k})\tau_z\psi_2({\bf
k})+\psi_2^\dagger({\bf k})\tau_z\psi_1({\bf k})),
\eeq
where we used Nambu notations:
$\psi_{\alpha,\uparrow}=c_{\alpha,\uparrow}$,
$\psi_{\alpha,\downarrow}=c_{\alpha,\downarrow}^\dagger$,
$\alpha=1,2$ corresponds to two different layers, and $\tau_x$ and
$\tau_z$ are the Pauli matrices. It is convenient to introduce
symmetric and antisymmetric combinations:
$\psi_+=(\psi_1+\psi_2)/\sqrt{2}$ and
$\psi_-=(\psi_1-\psi_2)/\sqrt{2}$. The the Hamiltonian splits into
the symmetric and antisymmetric parts $\mathcal H=\mathcal H_+ +
\mathcal H_-$:
\be
\mathcal H_{\pm}=\sum_{\bf k} \psi_\pm^\dagger({\bf k})
\left[(\varepsilon_{\bf k}\pm t_\perp)\tau_z+\Delta_{\bf
k}\tau_x\right]\psi_\pm ({\bf k}).
\ee

Let us now find the effect of $t_\perp$ on $A_2$ using
Eq.~(\ref{B_Q}). Reexpressing the operators $c_\alpha$ and
$c^\dagger_\alpha$ through the Nambu spinors $\psi$ and
$\psi^\dagger$ and using Wick's theorem and expanding to the leading
order in the tunneling coupling $t_\perp$ it is straightforward to
show that
\be
A_2=A_{2,1}+A_{2,2}+A_{2,3},
\ee
where
\beq
&&A_{2,1}\approx {\mathcal A\over 2}\int {d{\bf k}\over
(2\pi)^d}{\Delta_{\bf k}^2\over E_{\bf k}^2}\\
&&A_{2,2}\approx{\mathcal A^2 t_\perp^2\over 4}\left[ \int{d{\bf
k}\over (2\pi)^d}
{\Delta_{\bf k}^2\over E_{\bf k}^3}\right]^2,\\
&&A_{2,3}\approx 2\mathcal A t_\perp^2\int {d{\bf k}\over
(2\pi)^d}{\Delta_{\bf k}^4\over E_{\bf k}^6}.
\eeq
Here $E_{\bf k}^{\pm}=\sqrt{(\varepsilon_{\bf k}\pm
t_\perp)^2+\Delta_{\bf k}^2}$.

Note that $A_{2,1}$ coincides with our earlier expression
(\ref{B_Q4}). The two other terms $A_{2,2}$ and $A_{2,3}$ are
proportional to $t_\perp^2$. We point that $A_{2,2}$ scales faster
with the imaging area than the other terms. The reason is that even
in the absence of superfluidity the tunneling forces fermions to
occupy preferably the symmetric state. This is in turn equivalent to
establishing a well defined relative phase between the two atomic
systems. Since we are interested in detecting $A_{2,1}$ and not
$A_{2,2}$ in fermionic systems, contrary to bosonic, it is
preferable to make the imaging area as small as possible, of the
order of the Cooper pair size (or superconducting coherence length).
Indeed the integral in Eq.~(\ref{B_Q2}) converges at long distances
thus there is no need to go to ${\bf r}$ larger than the coherence
length. The third term $A_{2,3}$ comes from the Andreev process. It
is always subdominant at small $t_\perp$ and we can safely ignore
it.

For the $s-$ wave pairing the integrals above can be explicitly
evaluated:
\beq
&&A_{2,1}\approx {\mathcal A\rho_0\Delta\over 2}\arctan {\Delta\over\mu},\\
\label{B12}
&&A_{2,2}\approx \left[{\mathcal A t_\perp \rho_0\over 2}
\left(1+{\mu\over\sqrt{\mu^2+\Delta^2}}\right)\right]^2.
\label{B22}
\eeq
In the weak and strong pairing regimes we have the following
asymptotics:
\beq
A_{1,2}&\approx& { \mathcal A\rho_0\Delta^2\over
2\mu^2},\;A_2^2\approx
\mathcal A^2 \rho_0^2t_\perp^2,\quad{\rm \Delta\ll\mu}\label{25}\\
A_{1,2}&\approx& {\pi\over 4} \mathcal A\rho_0\Delta,\; A_2^2\approx
{1\over 2}A^2 t_\perp^2\rho_0^2,\quad{\rm \Delta\gg\mu}.
\label{26}
\eeq
Both at weak and strong pairing we find that
\be
{A_{2,2}\over A_{2,1}}\sim \sqrt{n\mathcal A}\, {t_\perp\over
\Delta},
\label{27}
\ee
where $n$ is the atom density. The $A_{2,2}$ contribution is an
unwanted correction coming from the interlayer coupling, which have
nothing to do with superfluidity. This contribution can be
suppressed either by decreasing the tunneling amplitude $t_\perp$ or
by decreasing the imaging area $\mathcal A$. Note that the tunneling
amplitude can not be pushed down all the way to zero, because then
one will loose phase coherence between the two layers. In the next
section we will see that it is always possible to find the regime
where $A_{2,2}$ is negligible and at the same time the two
superfluids are locked. We also comment that one can distinguish two
contributions $A_{2,1}$ and $A_{2,2}$ by looking into the dependence
of $A_2$ on $t_\perp$.

\subsection{Phase locking transition}

In this section we examine the effect of the tunneling coupling on
establishing phase coherence between the two fermionic superfluids.
From a prior work it is known that tow energy phase fluctuations in
superfluids can be described by means of a conventional $xy$ model
with the effective Lagrangian~\cite{simanek, thouless, stone}:
\be
\mathcal L=n \dot\theta-{n\over 4m}(\nabla\theta)^2-{1\over
16mn}(\nabla n)^2-{(n-n_0)^2\over 2\rho_0},
\label{lagr}
\ee
where
\be
n=n_0-\rho_0(\dot\theta+(\nabla\theta^2)/4m)
\ee
plays the role of the superfluid density, $\theta$ is the phase of
the superfluid order parameter. We emphasize that if fluctuations of
$\theta$ are small then $n$ is close to the total density of
atoms. It is straightforward to generalize the derivation of the
Lagrangian given in Ref.~[\onlinecite{thouless}] to the bilayer
system. A similar derivation  but for a tunneling through a point can be also found in Ref.~[\onlinecite{simanek}] and it is sketched in
Appendix B for completeness. We use this formalism to evaluate the
tunneling Hamiltonian from the effective action given in
Eq.~(\ref{effStun}).

In the case of $s-$ and $d$-wave pairings the integral appearing in
Eq.~(\ref{effStun}) can be easily evaluated yielding the following
coupling term to the Hamiltonian of the system:
\be\label{eff-tun}
\mathcal H_{12}=-\frac{t_\perp^2}{8\pi}\rho_0 T(\Delta/\mu)\cos
(\theta_1-\theta_2).
\ee
where the functions $T(\Delta/\mu)$ for the $s-$ and $d-$ wave
pairings are given in Appendix C. In both cases $T(\Delta/\mu)$ is a
smooth function, which interpolates between $T(\Delta/\mu)\approx 2$
at $\Delta\ll\mu$ and $T(\Delta/\mu)\approx 1$ in the opposite
limit.
When the pairing is small $\Delta\ll \mu$ the effective Josephson
coupling becomes simply $J\approx t_\perp^2\rho_0$, i.e. independent
on $\Delta$. This is a bit surprising result since one would naively
expect that $J$ should vanish as $\Delta\to 0$. And this is indeed
the case in superfluids with inhomogeneous tunneling where $J\sim
t_\perp^2 \rho_0^2\Delta$ (see Ref.~[\onlinecite{simanek}] for
details). In superfluids with uniform $t_\perp$ the Josephson
coupling gets enhanced by the coherence factor $1/(\rho_0
\Delta)\sim k_f\xi$, where $\xi$ is the BCS coherence length, and
the dependence of $J$ on $\Delta$ disappears.

The minimal tunneling required to lock the two phases together
between two superfluids can be estimated from equating the energy
gap required to transfer one particle from one superfluid to another
\be
E_c\approx {1\over \rho_0 L^2}
\ee
to the energy gain due to the tunneling term
\be
E_J\approx t_\perp^2 \rho_0 L^2,
\ee
where $L^2$ is the area of each condensate. Note that $L^2$ can be
significantly larger than the imaging area $\mathcal A$. From these
equations we find that $E_J>E_c$ is equivalent to
\be
t_\perp> {1\over \rho_0 L^2}.
\ee
This condition is compatible with the dominance of $A_1$ over $A_2$
in Eq.~(\ref{27}) if
\be
\Delta> {1\over \rho_0 L^2}\sqrt{n\mathcal A}.
\label{rec}
\ee
In the weak coupling limit $\Delta\ll \mu$ this requirement reduces
to $\Delta>\mu \sqrt{\mathcal A/n L^4}$ and in the strong coupling
regime $\Delta\geq \mu$ Eq.~(\ref{rec}) reduces to $\sqrt{\mathcal
A/nL^4}\ll 1$. Clearly both conditions can be easily satisfied using
either small imaging areas or systems with sufficiently large number
of particles $nL^2$.

For convenience we assumed the zero temperature case throughout the
paper. However, we would like to stress that our results will be
robust to the effects of temperature as long as it is below the
temperature corresponding to the Kosterlitz-Thouless transition.
Indeed in coupled condensates it is the global phase which is
destroyed by thermal fluctuations, but the relative phase remains
locked all the way to the Kosterlitz-Thouless
transition~\cite{mathey}.

\subsection{Discussion}
Having the effective action for the fluctuations of the order
parameter phase we can lift the assumption used in derivation of
Eq.~(\ref{B12}) that the phases between two superfluids are locked
together and get
\beq
&&A_{2,1}\approx {\mathcal A\rho_0\Delta\over 2}\arctan
{\Delta\over\mu}\mathrm e^{-\langle\theta^2\rangle/2}\approx
{\mathcal A\rho_0\Delta\over 2}\arctan {\Delta\over\mu}\nonumber\\
&&~~~~~~\times \exp\left(-{1\over 2\rho_0 t_\perp T_{S,D}(\Delta/\mu)L^2}\right).
\label{B13}
\eeq
We note that $A_{2,2}$ is not sensitive to the relative phase
between the two superfluids and thus Eq.~(\ref{B22})  holds for an
arbitrarily small $t_\perp$.
\begin{figure}[ht]
\includegraphics[width=8cm]{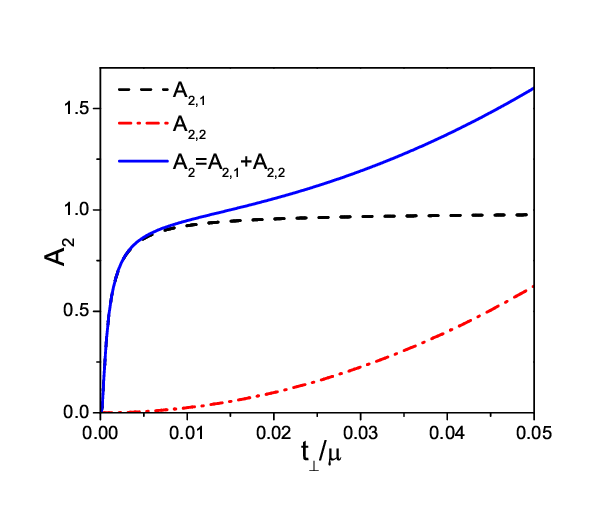}
\caption{Anomalous interference amplitude $A_2=A_{2,1}+A_{2,2}$  and
separately contributions $A_{2,1}$ (Eq.~(\ref{B13})) and $A_{2,2}$
(Eq.~(\ref{B22})) versus tunneling $t_\perp/\mu$ for sample
parameters (see text for details). The red line ($A_{2,2}$) is the
contribution, which comes from normal fermions, the black line is
the superfluid contribution.}
\label{pair_1}
\end{figure}

In Fig.~\ref{pair_1} we plot $A_2=A_{2,1}+A_{2,2}$  and separately
contributions $A_{2,1}$ (Eq.~(\ref{B13})) and $A_{2,2}$
(Eq.~(\ref{B22})) versus tunneling $t_\perp/\mu$ for sample
parameters $\Delta/\mu=0.2$, $\mathcal A/L^2=0.02$, $N=10^3$, where
$N$ is the total number of fermions per condensate. The graph
indicates that there is an intermediate tunneling regime, where the
superfluid contribution $A_{2,1}$ dominates over the normal part
$A_{2,2}$. We note that because $A_{2,2}$ is not sensitive to the
superfluid gap one can perform separate measurements of $A_2$ in the
normal and superfluid regimes. Then the difference between the two
will be precisely given by $A_{2,1}$.

So far considering interlayer coupling we focused only on large
imaging areas, with both transverse and longitudinal dimensions
large compared to microscopic length scales. In the opposite limit
one has to study the correction to Eq.~(\ref{B_Q2}). Instead of
repeating rather tedious analysis of Sec.~\ref{sec:cor} we will make
a couple of simple points. (i) There will be additional contribution
to $a_{2,2}(\theta)$ which scales as $t_\perp^2$. This contribution
will have the same origin as $A_{2,2}$ in Eq.~(\ref{B22}) and will
be insensitive to superfluidity. (ii) This unwanted contribution
will be greatly suppressed because it scales as the square of the
imaging area, which is small since one length scale is microscopic.
Thus the effect of the inter-layer coupling on the interference in
this case will be even smaller than in the case of the macroscopic
imaging area.

\section{Anomalous correlation functions in Bose systems}
\label{sec:3}

So far the main focus of our work was analysis of the possibility of
measuring anomalous correlation functions in fermionic superfluids.
There such measurements are key for determination of the pairing
gap. On the contrary one can get substantial information about the
superfluid properties of Bose systems analyzing normal correlation
functions~\cite{paris,PAD,GADP}. Nevertheless the possibility to
measure anomalous correlation functions can provide additional
valuable information about properties of these systems. As we will
see below these functions have very unusual behavior in the systems
with quasi-long order like zero-temperature one-dimensional and
finite-temperature two-dimensional systems. In this section we will
give explicit results both for 1D and 2D systems.

We consider setup analogous to that discussed in
Sec.~\ref{sec:anom}. Using the same arguments we find that
\be
A_2=\mathcal A\int_{{\bf r}\in\mathcal A} d^d r F_1^\star({\bf r})
F_2({\bf r}),
\ee
where $F_{\alpha}({\bf r})=\langle a_\alpha({\bf r}) a_\alpha (0))$,
$\alpha=1,2$; $\mathcal A$ is the imaging area for $2D$ systems and
the imaging length in the $1D$ setup. The operators $a_{\alpha}$
have bosonic statistics. As in the case with fermions $A_2$ vanishes
if the two systems are uncoupled. However, as we argue below if the
imaging area is smaller than the system size, one can always find
the regime of small transverse tunnelings $t_\perp$ such that the
phases of two superfluids are locked together, but the correlation
functions remain independent of $t_\perp$. In this case assuming
that the two systems are identical, as in Sec.~\ref{sec:anom}, we
have
\be
A_2\approx \mathcal A\int_{{\bf r}\in\mathcal A} d^d r |F({\bf
r})|^2,
\label{BQb}
\ee
where $F$ describes anomalous correlation functions in either of the
two systems. We note that the scheme I, where we use two imaging
areas can not be straightforwardly applied to bosons because the
single particle correlation functions decay slowly. However, in
bosonic systems one can deal even without tunneling for the
following reason. The relative phase between two independent
superfluids is random from shot to shot. Nevertheless in each shot
the interference amplitude is well defined and fluctuates only
weakly~\cite{PAD, paris}. Thus this unknown relative phase can be
reliably determined in each run. Then one can evaluate anomalous
correlation functions putting the origin of integration along $z$ in
the position of the central interference peak. It is easy to see
that this procedure eliminates the effect of the unknown phase and
leads to Eq.~(\ref{BQb}).

\subsection{One-dimensional superfluids}

Let us first analyze Eq.~(\ref{BQb}) for two coupled Bose systems at
zero temperature. If the imaging length is larger than the healing
length of the condensate then boson-boson correlation functions are
approximately
\be
\langle a(x)a(0) \rangle \propto \left(\frac{x}{L}\right)^{1\over
2K}+\ldots,
\ee
where $K$ is the Luttinger parameter related to the interaction
strength (see e.g. Ref.~[\onlinecite{caz}]) and $L$ is the length of
a system. Dots indicate other contributions which scales with larger
power and thus their contribution is less important. For
a Lieb-Liniger gas with short range repulsive interactions $K\gg 1$ in
the weakly interacting Gross-Pitaevskii regime and $K\to 1$ in the
fermionized Tonks-Girardeau regime. Notice that this correlation
function increases with the distance. This unusual behavior also
reflects the scaling of $B_Q^2$ with the imaging length $\mathcal
A$:
\beq
&&A_2(\mathcal A)\sim \tilde C\mathcal A^2 \left(\mathcal A\kappa
\right)^{1\over K},
\; \mathcal A\kappa\ll 1\nonumber\\
&&A_2(\mathcal A)\sim \tilde C\mathcal A^2,\quad \mathcal A\kappa\gg
1,
\eeq
where $\tilde C$ is a nonuniversal constant and
$\kappa\propto\sqrt{t_\perp}$ is the characteristic wavevector
corresponding to the transverse tunneling (see details below). At
sufficiently short distances the anomalous interference amplitude
$A_2$ grows faster than the area squared. As the imaging size
approaches the cutoff length the superlinear dependence becomes
linear and we recover the expected result for the system with a
long-range order.

We can make the analysis more quantitative using low-energy
description of two coupled one-dimensional Bose systems (see e.g.
Ref.~[\onlinecite{gpd}] for more details). In particular, for weakly
interacting superfluids it can be shown that the Lagrangian
governing properties of the relative phase $\phi(x,\tau)$ between
the condensates is:
\be
\mathcal L(x,\tau)={v_s K\over 4\pi}\left[{1\over
v_s^2}(\partial_\tau
\phi)^2+(\partial_x\phi)^2-2\kappa^2\cos(\phi)\right],
\label{lag}
\ee
where $\tau$ is the imaginary time, $v_s$ is the sound velocity in
each condensate, and $\kappa^2\approx 4\pi t_\perp n/v_s K$.

The analysis of either normal or {\it anomalous} correlation
functions can be performed using the form-factor approach. For more
details on this approach we refer to Ref.~[\onlinecite{EK}] for a
general scheme and  to Ref.~[\onlinecite{gpd}] for the application
to the one-dimensional condensates. This analysis is rather involved
for generic value of the interaction parameter $K$ and requires
summation of many contributions coming from the intermediate
processes including creation and annihilation of soliton-antisoliton
pairs as well as their bound states- breathers. Anomalous
correlation functions would correspond in that case to the
soliton-creating or soliton-annihilating form-factors \cite{LZ}. On
the other hand for weak interactions (large $K$) the Lagrangian can
be simplified even further if we invoke Gaussian approximation
replacing $\cos(\phi)$ by $1-\phi^2/2$. More carefully, this can be
done using the self-consistent harmonic approximation \cite{gia}.
In this simple Gaussian approximation the calculation of anomalous
correlation function becomes trivial and we get
\beq
&&F(x)=\langle a_1^\dagger(x) a_2(x) a_1^\dagger(0)
a_2(0)\rangle\approx C \langle \mathrm
e^{i(\phi(x)+\phi(0))}\rangle\nonumber\\
&& =C^\prime \left({\kappa\over\Lambda}\right)^{1\over
K}\!\exp\left(-{2\pi\over K}{1\over \kappa L}\right)\exp\left[-
{1\over K} K_0(\kappa x)\right]\!,
\label{fx}
\eeq
where $C$ and $C^\prime$ are nonuniversal numerical factors and
$\Lambda$ is the short-distance cutoff of the order of
inter-particle density $\Lambda\sim n$ (we assume that
$\kappa/\Lambda\ll 1$). The first exponential factor appearing in
the equation above is similar to that, which we discussed earlier
(see Eq.~(\ref{B13})). It shows that for small enough $t_\perp$ such
that $\kappa\lesssim 2\pi/KL$ the phases between two superfluids are
not locked and anomalous correlations are exponentially suppressed.
If on the other hand the opposite is true the phases are locked
together and this term is close to unity. And finally the last
multiplier gives the spatial dependence of the correlation function
$F(x)$. At $\kappa x\ll 1$ we have $K_0(\kappa x)\approx -\ln(\kappa
x)$ and we recover the asymptotic $F(x)\propto x^{2/K}$,in the
opposite limit $\kappa x\gg 1$ we have $K_0(\kappa x)\ll 1$ and thus
$F(x)\approx {\rm const}(x)$.

Using Eq.~(\ref{fx}) we find that
\beq
&&A_2 =\mathcal A C^{\prime\,2}
\left({\kappa\over\Lambda}\right)^{2\over K}\exp\left[-{4\pi\over
K}{1\over \kappa
L}\right]\nonumber\\
&&~~~~~~~\times\int\limits_{0}^{\mathcal A} dx \exp\left[- {2\over
K} K_0(\kappa x)\right].
\eeq

\begin{figure}[ht]
\includegraphics[width=8cm]{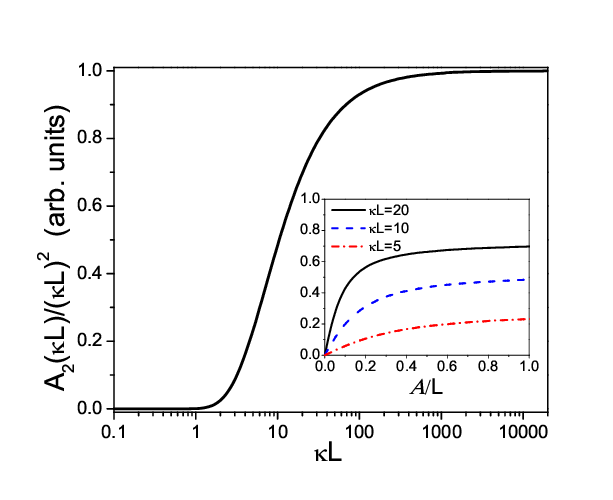}
\caption{The ratio of $A_2$ and $(\kappa L)^2$ vs. $\kappa L$ for
imaging length equal to the system size $\mathcal A=L$;
$\kappa\propto\sqrt{t_\perp}$ is the characteristic wavevector
related to the gap $\Delta$ (see Eq.~\ref{lag}). The inset shows
$A_2/\mathcal A^2$ as a function of $\mathcal A/L$ for different
values of $\kappa L$. The superlinear behavior of $A_Q$ vs.
$\mathcal A$ is a consequence of unusual behavior of anomalous
correlation function $F(x)\sim \langle \exp[i(\phi(x)+\phi(0))]$.}
\label{pair_1_2}
\end{figure}

In Fig.~\ref{pair_1_2} we plot the dependence of $A_2/(\kappa L)^2$
on $\kappa L$ for the imaging size equal to the system size
$\mathcal A=L$. At small tunneling $\kappa L\ll 1$ the two
condensates are decoupled and the anomalous correlator is
exponentially suppressed. As the transverse tunneling increases
$\kappa L\sim 1$ the anomalous interference amplitude increases
faster than linearly and at $\kappa L\gg 1$ we have $A_2\propto L$.
Note that there is a very wide range of parameters $3\lesssim\kappa
L\lesssim 100$, where the superlinear behavior can be observed.
Similarly one can fix the transverse tunneling and the system size
and analyze the dependence of $A_2$ as a function of the imaging
size. The inset shows such plots for different values of $\kappa L$.
Again one observes the superlinear behavior of $A_2$ in a wide
regime of parameters. We remind that the normal interference
amplitude $A_Q$ defined in Sec.~\ref{sec:norm} always has sublinear
behavior~\cite{PAD,GADP,IGD}.

We can easily generalize the analysis above to the case of finite
temperatures. In the regime when the thermal coherence length
$\xi_T\approx Kv_s/T$ is large compared to the healing length of the
condensates the phase model described the Lagrangian (\ref{lag})
gives the correct low energy description of the two coupled
superfluids. Within this model one finds
\be
\langle a(x) a(0)\rangle = n\exp\left[-{4\pi\over KL}\sum_q {\coth
{v_s\sqrt{q^2+\kappa^2}\over 2T}\over\sqrt{q^2+\kappa^2}}\cos^2 {q
x\over 2}\right].
\ee
In the zero temperature limit this expression reduces to
Eq.~(\ref{fx}) while at $T\geq v_s\kappa$ we get
\be
\langle a(x) a(0)\rangle\approx n\exp\left[-{\pi T\over K\kappa
v_s}\left(1+\mathrm e^{-\kappa x}\right)\right].
\ee
In this case we have exponentially increasing correlations
\be
\langle a(x) a(0)\rangle\sim \exp\left({\pi T (x-2/\kappa)\over K v_s}\right)
\ee
for $\kappa x\leq 1$ and then their saturation as $\kappa x$ becomes
larger than one. In turn this behavior of the correlation functions
implies that in the regime $T\geq \kappa v_s$ the anomalous
interference amplitude scales as
\be
A_2\propto \mathcal A\exp\left({2\pi T\over K v_s}\mathcal
(A-2/\kappa)\right)
\ee
at $\mathcal A\ll 1/\kappa$ and then in the usual way $A_2\sim
\mathcal A^2$ in the opposite limit. Thus anomalous correlation
functions can be used to probe the temperature in the system. We
note, however, that in the high temperature regime $T\gg \kappa v_s$
the anomalous interference amplitude $A_Q$ is exponentially
suppressed.

\subsection{Two-dimensional superfluids.}

In a similar fashion to the previous section we can analyze behavior
of the anomalous correlation functions and the anomalous
interference amplitude for two-dimensional bosonic superfluids.
Below the Kosterlits-Thouless phase transition temperature the low
energy properties of the superfluids can be described by the
effective Lagrangian (defined as the ratio of the energy density to
the temperature), which is very similar to (\ref{lag}):
\be
\mathcal L({\bf r})\approx {\hbar^2\rho_s\over 4 m
T}\left((\nabla\phi)^2-2\kappa^2\cos\phi\right),
\label{lag1}
\ee
where as before $\phi$ is the local phase difference between the two
superfluids, $T$ is the temperature, $\rho_s$ is the superfluid
density, $m$ is the boson's mass, and $\kappa^2\approx
4mt_\perp/\hbar^2$. Because of the formal analogy of Lagrangians
(\ref{lag1}) and (\ref{lag}) the anomalous correlation function
$F({\bf r})=\langle a({\bf r})a(0)\rangle$ in two dimensions is
identical to Eq.~(\ref{fx}) under the substitution $K \to
\pi\hbar^2\rho_s/(mT)$ and $v_s\to 1$. So one can expect a similar
superlinear behavior of the anomalous interference amplitude $A_2$:
\be
A_{2}\sim \mathcal A^{2+1/2K}
\ee
for $\kappa \sqrt{\mathcal A}\ll 1$ and $A_{2}\sim \mathcal A^{2}$
in the opposite limit.

\subsection{Paired multicomponent bosonic condensates}

Condensates of atom pairs can also be realized with bosonic atoms.
The original idea of fragmented condensates goes back to Nozieres
and Saint James\cite{nozieres}. They emphasized the difficulty of
achieving such states since attraction between bosonic atoms favors
binding not just two but many particles and may lead to the system
collapse. More recently, paired condensates were discussed by Kuklov
et al.~\cite{kuklov} for a two component bosonic mixture in an
optical lattice. Perhaps the most natural setting for the appearance
of bosonic pairing is polar condensates of $S=1$ atoms in two
dimensions\cite{ohmi,ho,zhou}. As discussed in
Refs.~[\onlinecite{mukerjee},\onlinecite{podolsky}], in such systems
general topological considerations suggest the appearance of
quasi-long range order for singlet pairs rather than individual
spinor components. This is the system that we focus on in this
subsection.

Let $\psi_\alpha(r)$ be individual spinor components with $m=\pm 1,
0$. We can make a spin singlet pair operator $\Delta(r) =
\psi_{+1}(r) \psi_{-1}(r)+ \psi^2_{0}(r)$. As discussed in
Refs.~[\onlinecite{mukerjee, podolsky}] for two dimensional polar
condensates, such as $^{23}Na$, one expects to find a phase in which
$\langle \Delta^\dagger(r_1) \Delta(r_2) \rangle$ shows power law
correlations. At the same time there are only short range
correlations for individual spin components.  In an interference
experiment from a pair of independent $S=1$ polar condensates one
should measure interference amplitude for individual spin
components, $A_{ m \, R}$, then construct $A_{R} = A_{+1\, R} A_{-1
\,R} + A_{0 R}$. In each shot the phase of $A(R)$ is random, so one
can again take two regions, $R_{I}$ and $R_{II}$, and consider
$\langle A_{R_{I}}^\dagger A_{R_{II}}\rangle $. This expectation
value should decay as a power law of the distance between the two
regions.

\section{Summary and Conclusions}
\label{sec:4}

In this paper we addressed questions of application of interference
experiments to detect paired states of either fermions or bosons in
low dimensions. We showed that direct generalizations of approaches
used in analyzing interference of independent bosonic condensates do
not work due to overwhelming shot noise contribution. Thus we
proposed and analyzed two alternative schemes of interference
experiments which can be used to study anomalous correlation
functions, which contain information about pairing amplitudes. These
(gauge-noninvariant) correlation functions provide complimentary
information to normal correlation functions and can be used to
characterize the properties of the superfluids.

It was shown how the method of studying the anomalous functions can
be used to detect various pairing orders. One of the scheme we
propose is based on the phase sensitive detection employed earlier
in the condensed matter systems. On the other hand, another scheme
deals with two superfluids weakly coupled by interlayer tunneling.
We establish the condition of validity of this scheme which involves
the tunneling strength, imaging area and the system size. We
emphasized important roles of different form of expansion
(transversal and longitudinal) and directions of observation. In the
case of bosonic superfluids anomalous correlation functions have an
unusual property that they increase with the separation between
quasiparticles.


\section{Acknowledgements}
We would like to thank  E. Altman, A. Aspect, J. Dalibard, V.
Galitski, M. Greiner, Z. Hadzibabic, M. Lukin, J. Schmiedmayer, and
M.~Zwierlein for discussions.  A.P. acknowledges support from AFOSR
YIP. V.G. and E. D. are supported by AFOSR, DARPA, MURI, NSF
DMR-0705472, Harvard-MIT CUA.

\appendix

\section{Expectation values of the normal correlations
for the systems with pairing} \label{App:A}

In the BCS approximation the
expectation value of the square of the interference fringe amplitude
discussed in Sec. II can be found as
\be
\langle |A|^2\rangle = \mathcal A\int {d{\bf k}\over (2\pi)^d}
\left(1-{\xi_{\bf k}\over E_{\bf k}}\right)^2 ,
\label{A_Q4}
\ee
where $n_f(E_{\bf k})$ is the Fermi distribution function, $E_{\bf
k}=\sqrt{\xi_{\bf k}^2+\Delta_{\bf k}^2}$, $\xi_{\bf k}={\bf
k}^{2}/2m -\mu$, $\mathcal A$ is the imaging area of the
interference. The pairing gap function $\Delta_{k}$ is a constant
$\Delta_{0}^{(s)}$ for the $s$-wave pairing and the $k$-dependent
function $\Delta_{k}=\Delta_{0}^{(d)}(\cos(k_{x})-\cos(k_{y}))$ for
the case of the $d$-wave pairing. In the absence of
superconductivity and at zero temperature $|A|^{2}=2N$, where $N$ is
the total number of fermions in each system. The nonzero value of
$|A|^{2}$ in this case simply reflects anti-bunching of fermions
(note the negative sign in Eq.~(\ref{A_Q})). As expected for
fermions $A\sim \sqrt{N}$ (see e.g. Ref.~[\onlinecite{P}]).

Considering $s$-wave pairing and using the constant density of
states $\rho_{0}$ in $2D$ one finds
\beq
|A|^{2}=\!2N\!-\!\mathcal A\Delta_{0}^{(s)}\rho_0\left({\pi\over
2}+\arctan(t_{s})-2[t^{2}_{s}+1]^{\frac{1}{2}}\right)\!\!,\nonumber\\
\label{A_Q2}
\eeq
where $t_{s}=\mu/\Delta_{0}^{(s)}$ and $\mu$ is the chemical
potential. In the BCS limit $\mu\gg \Delta_{0}^{(s)}$ the equation
above reduces to $A_2\approx 2N-\mathcal
A\pi\rho_0\Delta_{0}^{(s)}$. If we formally extrapolate $BCS$ theory
towards the unitarity limit $\mu\to 0$ and $\Delta_{0}^{(s)}\to
n/\rho_0$ then $A_2\to N(2-\pi/2)$. So we see that $A_2$ is a
monotonically decreasing function of the pairing gap. The results
will be somewhat different for the superfluids with the $d-$wave
pairing in which case  is
\beq
\langle |A|^2\rangle&=&2N+2{\cal
A}\rho_{0}\Delta_{0}^{(d)}([t^{2}_{d}+1]^{\frac{1}{2}}\\&+&2t{\bf
E}(-t^{-2}_{d})+2(1+t_{d}^{2})^{\frac{1}{2}}{\bf
E}(\frac{1}{1+t_{d}^{2}}))
\eeq
where $t_{d}=\mu/\Delta_{0}^{(d)}$.

\section{Transversal vs. Longitudinal expansions}

In this Appendix we compare the two regimes of expansions: the
transversal expansion advocated in the main text versus longitudinal
one which is more close to to the standard time-of-flight technique.

The quantity similar to $A_2$ in Eq.~(\ref{B_Q2}) can be also
measured in the standard time of flights experiments. For the
low-dimensional superfluidity it is advantageous to have
longitudinal expansion so that the atoms from different layers do
not mix with each other. If one assumes that the interactions are
not important during the expansion then the spatial image of the
cloud after time of flight gives the momentum distribution of atoms
in the initial condensate. As it is shown in Ref.~[\onlinecite{adl}]
the density-density correlation maps to the Fourier transform of the
pairing amplitude:
\be
\langle n({\bf R},t)n(-{\bf R},t)\rangle=|F({\bf Q})^2|,
\label{nc}
\ee
where $Q=mR/\hbar t$. We note that the transverse expansion
discussed in the previous section directly probes the spatial
structure of (the square of) the pairing wave function and thus
gives a complimentary information to the quantity (\ref{nc}). If one
integrates Eq.~(\ref{nc}) over the momentum ${\bf Q}$ then one
recovers the expression identical to Eq.~(\ref{B_Q3}). So the two
setups give equivalent information. We note, however, that in the
transverse expansion regime one can benefit from independently
averaging over several imaging areas within a single shot.

Doing the longitudinal expansion one can also determine the spatial
structure of the pairing amplitude. Indeed if one integrates
Eq.~(\ref{nc}) along a preferred direction then the outcome should
be isotropic for the $s-$ wave pairing and highly unisotropic in the
$d-$ wave regime. Thus within the BCS model in the $s-$wave case and
in the $d-$ wave case if one integrates along the anti-nodal
direction (where the pairing gap is maximal) one gets
\be
\int dx\langle n({\bf R},t)n(-{\bf R},t)\rangle\propto \Delta
{\sqrt{\mu+\sqrt{\mu^2+\Delta^2}}\over \sqrt{\mu^2+\Delta^2}}.
\ee
On the other hand if one integrates Eq.~(\ref{nc}) along the nodal
direction in the $d-$ wave case (i.e. the direction where the
pairing gap vanishes) one should get zero.

The clear advantage of the longitudinal expansion method is that one
avoids the issues with coupling between different layers. In
principle, one can perform this experiment even on a single layer.
However, there is a big disadvantage too. Namely, one has to rely on
the free expansion to get the right correlation functions. In the
case of a bilayer system predominantly transverse expansion is
guaranteed by the large kinetic energy of the transverse
confinement. On the other hand for the longitudinal expansion
collisions between atoms shortly after expansion started can affect
the outcome of the experiment. We note that instead of purely
longitudinal expansion one can have the full three-dimensional time
of flight experiment and get similar results. However, if one is
interested in two-dimensional superfluidity allowing expansion in
all three dimensions one clearly looses in the signal to noise
ratio.

\section{Effective action approach to coupled systems}
The effective action ($S_{\rm eff}=\int_0^\infty \mathcal
L(\tau)d\tau$) for the phase fluctuations reads~\cite{thouless}:
\beq
&&S_{\rm eff}[\theta_1,\theta_2]={\rm Tr}\sum_{n=1}^\infty (\mathcal
G_0 \Sigma)^n\nonumber\\
&&-{1\over 2g}\int d^d r d\tau\,( |\Delta_1({\bf
r},\tau)|^2+|\Delta_2({\bf r},\tau)|^2),
\label{seff}
\eeq
where $g$ is the interaction strength, which we assume to be short
range, $\mathcal G_0$ and $\Sigma$ are $4\times 4$ matrices:
\be
\mathcal G_0=\left(\begin{array}{cc} G_{1} & 0\\
0 & G_{2}
\end{array}\right),\quad \Sigma=\left(\begin{array}{cc} \Sigma_1 & T\\
T^\dagger & \Sigma_2
\end{array}\right).
\ee
Here $G_{1,2}$ are the fermion's Green's functions in the superfluid
with a non-fluctuating phase. In the Nambu notation their inverses
are:
\be
G_{1,2}^{-1}=\left(\begin{array}{cc} \partial_\tau-
{\hbar^2\over 2m}\nabla^2 & |\Delta_{1,2}|\\
|\Delta_{1,2}| & \partial_\tau+{\hbar^2\over 2m}\nabla^2
\end{array}\right),
\ee
$\Sigma_{1,2}$ contains fluctuations of the order
parameter~\cite{thouless}:
\beq
&&\Sigma_{1,2}=-{\hbar^2\over
4m}(\nabla^2\theta_{1,2}+2\nabla\theta_{1,2}\nabla)\tau_0
\nonumber\\
&&+\left({\hbar\over 2}\dot\theta_{1,2}+{\hbar^2\over
8m}(\nabla\theta_{1,2})^2\right)\tau_z-\delta|\Delta_{1,2}|\tau_x.
\eeq
And finally $T$ corresponds to the tunneling coupling between two
layers:
\be
T=t_\perp\mathrm e^{i(\theta_1/2-\theta_2/2)\tau_z}\tau_z.
\ee
For simplicity we assume that two layers are identical and
$|\Delta_1|=|\Delta_2|\equiv \Delta$

Next we expand the effective action (\ref{seff}) to the second order
in small fluctuations in derivatives of $\theta_{1,2}$ and to the
second order in $t_\perp$. For simplicity we ignore fluctuations in
the magnitude of the order parameter $\delta|\Delta_{1,2}|$. Then
one obtains:
\be
S_{\rm eff}\approx S_{1}+S_2+S_{1,2},
\ee
where $S_{1,2}$ correspond to the quadratic Lagrangians (\ref{lagr})
of decoupled layers and
\be
S_{12}=-{\rm Tr}\, (G_0 T G_0 T^\dagger).
\label{s12}
\ee
Note that we can ignore slow spatial variations of the phases
$\theta_{1,2}$ in the tunneling matrix $T$ in the equation above.
Indeed keeping these variations will result to corrections to the
gradient terms in $S_{1,2}$ proportional to $t_\perp^2$. Then taking
into account the expression for the non-diagonal elements of the
Green function,
\beq
\hat{G}_{0}=-\frac{i\omega\tau_{0}+\xi_{k}\tau_{3}-\Delta_{k}
\tau_{1}}{\omega_{n}^{2}+\xi^{2}_{k}+\Delta_{k}^{2}}
\eeq
where $\xi_{k}$ is dispersion and $\tau_{0,1,3}$ are Pauli matrices
in Nambu space, Eq.~(\ref{s12}) can be evaluated in the limit of
zero temperature as
\beq\label{effStun}
S_{12}&=&-2 t_\perp^2 \rho_{0}\nonumber\\
&\times&\cos (\theta_1-\theta_2)\,\int {d\omega\over 2\pi}\int {d^2
k\over (2\pi)^2} {\Delta_{\bf k}^2\over (\omega^{2}+E_{\bf
k}^{2})^2}.
\eeq
where $n_{F}(0)$ is a density of state at the Fermi level and
$E_{{\bf k}}^{2}=\xi_{{\bf k}}^{2}+\Delta_{{\bf k}}^{2}$. For the
$s$-wave superfluid we use $\Delta_{k}\equiv\Delta$ and
$\xi_{k}=k^{2}/2m-\mu$ whereas for the $d$-wave pairing
$\Delta_{k}=\Delta_{0}(\cos(k_{x})-\cos(k_{y}))$.

\section{Technical details}

 In this Appendix we present some details of our
computations of contributions from the anomalous correlation
functions.  While the expression for the most functions in the case
of $s$-wave pairing can be evaluated straightforwardly, its
treatment for the $d$-wave case require some approximations which we
describe here. However, some integrals can be done without these
approximations which is also discussed.

\subsection{Nodal approximation}
In the $T\rightarrow 0$ limit we are interested in, to evaluate the
$d$-wave related integrals we use the nodal approximation developed
in \cite{PALee}. It consists in focusing on the regions close to
four nodes of the $d$-wave order parameter, ${\bf
k}_{n}=k_{F}\hat{k}_{n}$, ($n=1,2,3,4$) on the Fermi surface.
Explicitly,
\beq
{\bf\hat{k}}_{1}&=&\frac{{\bf\hat{x}}+{\bf\hat{y}}}{\sqrt{2}},\quad
{\bf\hat{k}}_{2}=\frac{-{\bf\hat{x}}+{\bf\hat{y}}}{\sqrt{2}},\\
{\bf\hat{k}}_{3}&=&\frac{-{\bf\hat{x}}-{\bf\hat{y}}}{\sqrt{2}},\quad
{\bf\hat{k}}_{4}=\frac{{\bf\hat{x}}-{\bf\hat{y}}}{\sqrt{2}}
\eeq
In the vicinity of nodes we can expand ${\bf k}={\bf
k}_{n}+\delta{\bf k}$, and then
\beq
\xi_{k}&=&v_{F}\delta k_{\perp},\quad \Delta_{{\bf
k}}=v_{\Delta}\delta k_{\parallel},\\
E_{{\bf k}}^{2}&=&v_{F}^{2}\delta k_{\perp}^{2}+v_{\Delta}^{2}\delta
k_{\parallel}^{2}
\eeq
Here $\delta k_{\perp,\parallel}$ are components of momentum
perpendicular and parallel to the Fermi surface, $v_{F}$ is a Fermi
velocity and $v_{\Delta}=|\partial\Delta_{{\bf k}}/\partial{\bf
k}_{\parallel}|$. Finally introducing the angular parametrization
\beq
\xi_{k}&\equiv& v_{F}\delta k_{\perp}=\epsilon\cos\alpha,\quad
\Delta_{{k}}\equiv v_{\Delta}\delta
k_{\parallel}=\epsilon\sin\alpha,\\ E_{{\bf k}}&=&\epsilon
\eeq
The integrals can be performed according to the following rule:
\beq
\int\frac{d^{2}{\bf
k}}{(2\pi)^{2}}(...)\rightarrow\sum_{n=1}^{4}\int\frac{d\delta
k_{\perp}d\delta k_{\parallel}}{(2\pi)^{2}}(...)\\
=\frac{1}{2\pi
v_{F}v_{\Delta}}\sum_{n=1}^{4}\int_{0}^{\epsilon_{max}}d\epsilon
\epsilon\int_{0}^{2\pi}\frac{d\alpha}{2\pi}(...)
\eeq
where the limit $\epsilon_{max}\simeq\Delta_{0}$ makes sure that the
integration area is equal to the area of the Brillouin zone. It can
be send to infinity for practical purposes.

Considering the tunneling integral of Eq.~(\ref{effStun}) we observe
that
\beq
\int_{0}^{\omega_{max}}\frac{d\omega}{2\pi}\int_{0}^{k_{max}} \frac{kdk}{(2\pi)^{2}} \\
=\frac{\epsilon_{max}}{2\pi^{2}
v_{F}v_{\Delta}}\left[\arctan(y)-y\log\left(\frac{y^{2}}{1+y^{2}}\right)\right]
\eeq
where $y\equiv\omega_{max}/\epsilon_{max}$. In the limit
$y\rightarrow\infty$ the expression in square brackets goes to $\pi$
and we arrive to the equation (\ref{eff-tun}) if one assumes that
the ratio $\epsilon_{max}/v_{F}v_{\Delta}$ is a constant independent
of $\Delta$.

\subsection{Explicit evaluation of some integrals}
\begin{figure}[h]
\includegraphics[width=8cm]{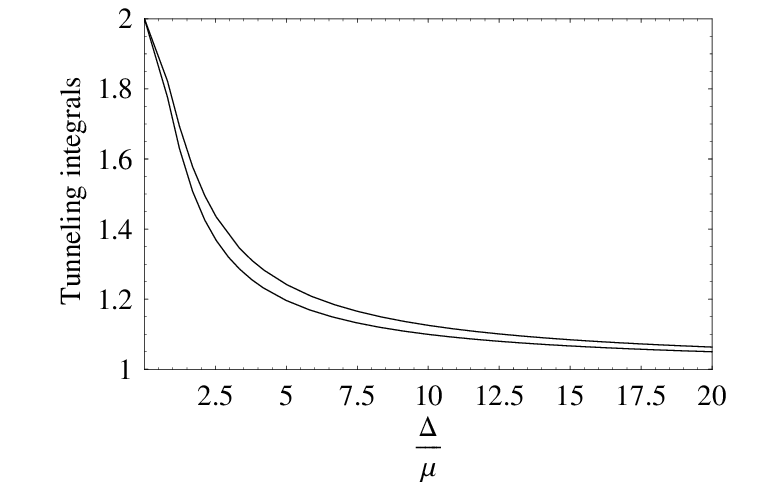}
\caption{Comparison of the tunneling elements for the $s$-wave
pairing (lower curve) vs $d$-wave pairing (upper curve) as a
function of }
\label{sw-vs-dw}
\end{figure}

Some explicit expressions are possible to obtain for several
important quantities: in particular we focus on the function $b_{Q}$
introduced in Sec.~\ref{probing-pairing} and the tunneling
integrals.

Starting from the expression for the anomalous Green function
\beq
F(r)=\int_{0}^{\infty}\frac{2m^{*}k
dk}{(2\pi)^{2}\hbar^{2}}\int_{0}^{2\pi}
\frac{e^{ikr\cos\phi(\frac{2m^{*}}{\hbar^{2}})^{\frac{1}{2}}}\Delta\cos2\phi
d\phi}{\sqrt{(k^{2}-\mu)^{2}+\Delta^{2}\cos^{2}2\phi}}\nonumber\\
\eeq
writing $\cos^{2}2\phi=(1/2)+\cos(4\phi)/2$ and using the expansion
formula
\beq\label{expand}
\frac{1}{\sqrt{1+x}}=\sum_{k=0}^{\infty}\frac{(-1)^{k}(2k)!x^{k}}{(k!)^{2}2^{2k}}
\eeq
for $x=\Delta^{2}\cos(4\phi)/2$ one can obtain a systematic
expansion of the integrand. Restriction to the first term in this
expression already produces a very good uniform approximation for
the integral. Doing the $\phi$ integration
\beq
\int_{0}^{2\pi}e^{ik\tilde{r}\cos(\phi)}\cos(2\phi)=-2\pi J_{2}(k
\tilde{r})
\eeq
where $\tilde{r}=r(2m^{*}/\hbar^{2})^{\frac{1}{2}}$ gives an
expression
\beq\label{dint}
F(r)\approx-\rho_{0}\Delta\int_{0}^{\infty} k dk \frac{
J_{2}(k\tilde{r})}{\sqrt{(k^{2}-\mu)^{2}+\frac{\Delta^{2}}{2}}},
\eeq
where $\rho_{0}=m^{*}/\pi\hbar^{2}$ is the constant density of
states in 2D (summed over spin polarizations). The expression
(\ref{dint}) can be further evaluated using the so-called
$\xi$-approximation, frequently used in the theory of
superconductivity. In the argument of the Bessel function we write
$k^{2}-\mu=(k-\sqrt{\mu})(k+\sqrt{\mu})\approx(k-\sqrt{\mu})2\sqrt{\mu}$.
On the other hand, in the BCS regime where $\mu/\Delta>>1$ after
variable's rescaling we shift the  lower limit of integration to
$-\infty$ and use the summation theorem for the Bessel function
$J_{\nu}(z+t)=\sum_{n=-\infty}^{\infty}J_{\nu-k}(t)J_{k}(z)$.
Restricting to the $n=0$ from this sum already produce a good
approximation for the function $F(r)$,
\beq
F(r)=
-\frac{\rho_{0}\Delta}{2}J_{2}(\sqrt{\mu}\tilde{r})I_{0}(\frac{\tilde{r}\Delta}{4\sqrt{2\mu}})K_{0}(\frac{\tilde{r}\Delta}{4\sqrt{2\mu}})
\eeq
Now, the integral for $b_{Q}^{2}$ can be computed in the same BCS
limit by using the asymptotic behavior of Bessel functions for large
and small arguments. The $\log^{3}$-result is similar to the
$s$-wave case but with different prefactor given in Eq.~(\ref{bQd}).

Tunneling integrals in the $d$-wave case can be evaluated exactly.
The answer is shown in Eq.~(\ref{eff-tun}) with the function
\beq
T_{D}(y)&=&1+\frac{2}{\pi y^{2}}[\sqrt{1+y^{2}}{\bf
E}(\frac{y^{2}}{1+y^{2}})\\
&+&{\bf E}(-y^{2})-{\bf K}(-y^{2})-\frac{1}{\sqrt{1+y^{2}}}{\bf
K}(\frac{y^{2}}{1+y^{2}})]\nonumber
\eeq
where ${\bf E}(y)$ and ${\bf K}(y)$ are complete elliptic integrals
and $y\equiv \Delta/\mu$. In Fig.~(\ref{sw-vs-dw}) it is compared
with exact expression for the $s$-wave tunneling integral
\beq
T_{S}(y)=1+\frac{1}{\sqrt{y^{2}+1}}.
\eeq
The similarity of the tunneling integrals for the $s-$ and $d-$ wave
pairings is remarkable.

\end{document}